\documentclass[11pt,dvips]{article}

\usepackage{amsmath,amsfonts,amssymb} 
\usepackage{amsthm}
\usepackage{graphicx}                 
\usepackage[round]{natbib}
\usepackage{geometry}
\usepackage{color}

\DeclareMathOperator{\diag}{\text{diag}}

\DeclareMathOperator{\eps}{\varepsilon}

\title{Integrative clustering of high-dimensional data with joint and individual clusters, with an application to the Metabric study}
\author{Kristoffer Hellton, Magne Thoresen \\
Department of Biostatistics, University of Oslo, \\
P.O.Box 1122 Blindern N-0317, Oslo, Norway \\
\emph{k.h.hellton@medisin.uio.no}}

\begin{document}
\maketitle

\begin{abstract}
When measuring a range of different genomic, epigenomic, transcriptomic and other variables, an integrative approach to analysis can strengthen inference and give new insights. This is also the case when clustering patient samples, and several integrative cluster procedures have been proposed. Common for these methodologies is the restriction of a joint cluster structure, which is equal for all data layers.  We instead present Joint and Individual Clustering (JIC), which estimates both joint and data type-specific clusters simultaneously, as an extension of the JIVE algorithm \citep{lock2013joint}. The method is compared to iCluster, another integrative clustering method, and simulations show that JIC is clearly advantageous when both individual and joint clusters are present. The method is used to cluster patients in the Metabric study, integrating gene expression data and copy number aberrations (CNA). The analysis suggests a division into three joint clusters common for both data types and seven independent clusters specific for CNA. Both the joint and CNA-specific clusters are significantly different with respect to survival, also when adjusting for age and treatment.
\end{abstract}

\noindent \textit{Keywords:} Breast cancer; Clustering; Integrative genomics; Latent variable estimation; Singular value decomposition.

\section{Introduction}
The rapid development in genomic technologies has enabled the analysis of an increasing range of data layers or data types. This increases the need for integrative procedures that can handle several data types. When studying diseases that build on several molecular processes, we need to consider the interplay between the genomic layers to fully understand the phenotypic traits. We should therefore attempt to integrate different data types in a single joint analysis, and this is the core principle of integrative genomics.  As the information content is higher in an integrative framework compared to individual analyses, it is possible to gain statistical power to detect relevant signals. This is especially relevant for genetically driven diseases such as cancer in general or breast cancer, as studied in this paper. 

An integrative approach is especially relevant in the exploratory field of unsupervised clustering, and such procedures have been suggested earlier \citep{shen2009integrative,shen2013sparse,lock2013bayesian}. The aim of clustering is to discover novel disease subtypes, which can aid the understanding of survival and mortality risk differences or enable personalized treatments. Earlier integrative clustering approaches include the iCluster methodology \citep{shen2009integrative,shen2013sparse} and the Bayesian consensus clustering \citep{lock2013bayesian}. The iCluster method clusters observations based on joint latent variables, utilizing the connection between k-means clustering and latent factor modeling. In Bayesian consensus clustering, observations are clustered for each data type separately with a last step of combining the different groupings into a consensus solution. 

However, when several highly heterogeneous genomic data types are integrated, some cluster structures are typically not shared between all the data layers. If there are clear clusters present in some of the data types, but not in others, these can confound or obscure the joint clusters shared by all data types. Data type-specific cluster structures can be caused by biological confounders, such as ethnicity, or technical and measurement-related differences, such as samples processed at different labs or changes in techniques over time, affecting only a single data type. But more importantly from a biomedical point of view, there could exist disease-related patient clusters that are independent of the joint subtypes, but still relevant and interesting for treatment and disease-understanding. 

Our aim is to take into account the presence of data type-specific clusters together with joint clusters in an integrative framework. We will therefore present a clustering extension of the JIVE algorithm \citep{lock2013joint}, which decomposes several data sets into joint and individual latent structures in an iterative procedure. In our extension, termed Joint and Individual Clustering (JIC), the joint cluster structure is estimated simultaneously with the individual or data type-specific clustering. JIC will be compared to the iCluster methodology in different simulation settings and will be used to find joint and data type-specific clusters of patients in the Metabric study \citep{curtis2012genomic}.

\section{Integrative clustering} 
The iCluster method \citep{shen2009integrative,shen2013sparse} has become an established method for integrative clustering of multiple genomic data types. We extend the JIVE methodology \citep{lock2013joint} to accommodate clustering of observations, as done by iCluster. Both approaches are based on estimating latent variables as continuous representations of the cluster assignment vectors. An important difference between JIC or JIVE and iCluster is the assumed noise structure in the latent variable model. iCluster allows the factor residuals to have different variances for each variable, while JIC, assuming equal variance,
allows for additional latent variables specific for each data type. Both approaches can incorporate sparsity in the loadings matrices. 

Integrative clustering aims to cluster observations simultaneously in different data types. Let $X_{1}, \dots, X_{M}$ be $M$ different genome-scale data types (typically expression, copy number variation, methylation) or genome-related data types (such as miRNA, proteins, transcription factors) that are all measured on the same $n$ patients, indexed $j=1,\dots,n$. Then each $X_{m}$ is a $p_{m} \times n$ data matrix for $m=1, \dots, M$ with $p_{m}$ variables, indexed by $i=1,\dots,p_{m}$. The data types can be highly heterogeneous with respect to scale, unit or variation.

The $M$ data matrices can be combined into a single concatenated matrix 
\[X = \begin{bmatrix} X_{1} \\ \vdots \\ X_{M} \end{bmatrix},\]
of dimension $p \times n$ where $p = p_{1}+\dots+p_{M}$. A scaled version of the concatenated matrix can be constructed by first scaling each data matrix $X_{i}$ by some norm $\|X_{i}\|$. Then each data type will contribute equally to the integrative solution. 

\subsection{Clustering and dimension reduction}
Both iCluster and JIC are closely linked to k-means clustering, where clusters are defined by minimizing the distance between each observation and the cluster centroid. To simplify the procedure of k-means clustering, one can use principal component analysis (PCA) as an initial step to reduce the dimension of the data matrix. This two-step procedure, called ``tandem clustering'' \citep{arabie1996advances,terada2014strong}, clusters the reduced subset of PC scores, but have been criticized in the statistics literature. 

However, in machine learning, \citet{zha2001spectral,ding2004k} have shown that principal components are the continuous solution to the k-means optimization problem, such that the PC scores correspond to a continuous version of the discrete cluster indicators. Specifically, if the k-means clustering solution is denoted $Z^{T}=[z_{1},\dots,z_{K-1}]$, a matrix of $K-1$ indicator vectors
\[z_{k}^{T} = n_{k}^{-1/2}[0,\dots,0,\underbrace{1,\dots,1}_{n_{k}},0,\dots,0],\]
where $n_{k}$ is the number of observations in each cluster, the $K-1$ first principal component scores will minimize the k-means objective function. Therefore, k-means clustering (into $K$ groups) can be solved in two steps: first find the $K-1$ (standardized) principal component scores, and then reconstruct the discrete cluster assignments from the continuous scores, for instance with k-means clustering. In a high-dimensional setting, this is highly efficient as the data matrix is reduced from $p \times n$ to $(K-1) \times n$. 

The estimation of the continuous matrix $Z$ can also be done through Gaussian latent variable modeling, where the data matrix $X_{m}$ is modeled as
\[X_{m} = W_{m}^T Z + \eps_{m},\quad \eps_{m} \sim N(0,\Sigma),\]
where $W_{m}$ is a loading coefficient matrix and $\eps_{m}$ is a set of independently distributed errors. \citet{tipping1999probabilistic} connected the latent factor model and PCA, showing that under homogeneous and normally distributed errors, $\Sigma =\sigma^2 I_{p_{m}}$, the maximum likelihood estimates of $W_{m}$  yield the same solution as classical principal component analysis. The use of latent variable modeling as a part of the k-means clustering is motivated by the natural extension of the latent variables to multiple data types. 

\subsection{iCluster}
The iCluster method extends k-means clustering to an integrative clustering procedure, following the same approach as \citet{van2009structured,van2011flexible}. The latent variables $Z$, representing the clusters, are assumed to be common for all the data types. iCluster assumes the following model for $M$ data types:
\begin{align*}
X_{1} &= W_{1}^{T}Z + \eps_{1}, \\
& \quad\vdots \\
X_{M} &= W_{M}^{T}Z + \eps_{M}, 
\end{align*}
where the noise terms are heterogeneous, $\eps_{m}\sim N(0, \Psi_{m}), \Psi_{m} =\diag(\sigma_{1}^2,\dots, \sigma_{p_{m}}^2)$. The parameter estimates are obtained by maximum likelihood estimation using the EM algorithm. If $\eps_{m}$ was homogeneous, the solution is analytically given by the singular value decomposition. In iCluster, one can also enforce sparsity on the loading matrices by penalizing the data log-likelihood. After convergence of the EM algorithm, the rows of $Z$ are clustered by the k-means algorithm to obtain the group membership of each observation. In this way, the latent variable $Z$ corresponds to a cluster indicator matrix shared between all data sets. 

\subsection{Joint and Individual Clustering (JIC)}
Clustering based on estimated latent variables can also include other noise structures. We present a novel clustering extension of JIVE, the Joint and Individual Clustering (JIC), where clustering is carried out on both joint and data type-specific latent variables. The JIVE scheme proposed by \citet{lock2013joint} decomposes multiple data matrices into joint and individual structures. Both the shared and the data type-specific latent variables can be used to obtain a clustering of patients in a finale reduced k-means step. 

In JIC, the data types are assumed to be realizations of a combination of common and data type-specific latent variables 
\begin{align*}
X_{1} &= W_{1}^{T}Z + V_{1}^{T}Z_{1} +\eps_{1}, \\
& \quad\vdots \\
X_{M} &= W_{M}^{T}Z + V_{M}^{T}Z_{M} + \eps_{M}, 
\end{align*}
where $\eps_{m}\sim N(0, \sigma_{m}^{2}I),m=1, \dots,M$ and the joint loading matrices form a concatenated matrix
\[W = \begin{bmatrix} W_{1}^{T} \\ \vdots \\ W_{M}^{T} \end{bmatrix}. \]
When each individual latent clustering matrix $Z_{m}$, is orthogonal to the joint latent matrix, such that $ZZ_{m}^T= 0_{(K-1)\times (K_{m}-1)}$, there exists a unique decomposition of $X$ \citep[Supplementary material]{lock2013joint}. The decomposition can be found by minimizing the reconstruction error
\[\|R\|^{2} = \sum_{m=1}^{M}\|R_{m}\|^{2}= \sum_{m=1}^{M} \|X_{m} -W_{m}^{T}Z - V_{m}^{T}Z_{m}\|^{2}. \] If the rank  of $W^{T}Z$, $r$, and the rank of $V_{m}^{T}Z_{m}$, $r_{m}$,  for $m=1,\dots M$ are fixed, the decomposition can be found by iteratively estimating the joint and individual structures: First fix $W^{T}Z$ and estimate each $V_{m}^{T}Z_{m}$ by minimizing $\|R_{m}\|$. Then fix $V_{1}^{T}Z_{1}, \dots, V_{M}^{T}Z_{M}$ and estimate $W_{m}^{T}Z $ by minimizing $\|R\|$. This procedure is repeated until a suitable convergence criterion is reached.

When errors are assumed homogenous across variables (of same type), the solution minimizing the reconstruction error is given by the singular value decomposition and the latent variables corresponds to the  left singular vectors or standardized principal component scores estimated as follows: 
\begin{itemize}
	\item Calculate $W^{T}Z$ by the $r$ rank singular value decomposition of $X$, and subtract $W^{T}Z$ from $X$,
	\item Calculate $V_{m}^{T}Z_{m}$ by the $r_{m}$ rank singular value decomposition of the sub-matrix $X_{m}-W_{m}^{T}Z$, for $m=1,\dots M$
	\item Form the concatenated matrix of $X^{(l+1)}_{m} = X^{(l)}_{m}-V_{m}^{T}Z_{m}$ for $m=1,\dots M$ and repeat all steps until convergence.
\end{itemize} 
At convergence, the rows of $Z^{T}$ are clustered into $r+1$ groups and the rows of $Z_{m}^{T}$ are clustered into $r_{m}+1$ groups for $m=1,\dots M$, respectively, using k-means clustering.  

%
%

\subsection{Procedure for selection number of clusters}
To choose the number of clusters is a difficult task, and in general there is no optimal procedure. However, the selection procedure can be tailored to the method and relevant data, and we will use a procedure enlightening the subjective choices always present in such analyses.

Firstly, we exploit the subspace structure in JIC. The number of dimensions present in the clustering step is directly given by the number of clusters we aim to find; for $K$ clusters, we use $K-1$ component scores. As these are given by the singular value decomposition, the variables are by construction uncorrelated with each other, $ZZ^{T} = I_{K-1}$, such that each dimension contains independent information regarding the clustering. As shown by \citet{ding2004k}, a new cluster should be separated out in each dimension specified by a component. We exploit this property, and check if a new cluster is present in each added dimension. When no new cluster separates out, the total number of relevant dimensions is found. We use the following procedure:  
\begin{enumerate}
	\item For the $i$th component, check if the k-means clustering into two clusters is better than one cluster by a chosen procedure. 
	\item If two clusters are better, proceed to the next component. If instead only one cluster is supported, stop and set the number of clusters to the current component number.
\end{enumerate}
Instead of checking $K$ clusters in a $K-1$ dimensional space, we will check two clusters in a one-dimensional space, until we find the first component where no new cluster is present. 

How to check the presences of a new cluster should depend on the application and data characteristics. Some possible choices of procedures are: 
\begin{itemize}
	\item \emph{Prediction strength \citep{tibshirani2005cluster, shen2013sparse}:} evaluates clusters based on reproducibility between random splits of  the data into discovery and validation sets. A predicted and validation clustering are evaluated by a similarity index, and the $K$ with the highest index is chosen. However, in the $p\gg n$ setting, the component scores are very stable \citep{lee2014convergence,hellton2014asymptotic}, such that the sub-sampling induces little variability. Therefore component scores representing noise can exhibit very good cluster reproducibility, a property which is not desirable. 
	\item  \emph{Cluster separation}: clusters can be evaluated by a separation criterion, such as the Calinski-Harabasz, the Dunn criterion or within group sum-of-squares. This requires the index value for a single cluster, which can be difficult to assess. The approach seems to work best in low-dimensional settings with well-separated clusters \citep{milligan1987methodology}.
	\item \emph{Approach of G-means \citep{hamerly2003learning}:} evaluates the normality of the continuous scores. 
When no clusters are present, the component scores should behave as noise and follow a normal distribution, instead of a mixing distribution. We can evaluate this normality by qq-plots or normality tests. If the scores deviate significantly from normality, they do not resemble pure noise and clusters are present in the data. If the test is not significant, there is no evidence of clusters beyond the normally distributed noise. This approach seems to work well when clusters are not well-separated, and instead resemble a continuum.
\end{itemize}

\subsection{Cluster procedure for JIC} \label{findingK}
As genetic data usually do not exhibit well-separated clusters, we will utilize the idea behind the G-means method together with the notion of the independent subspaces. We use qq-plots, complemented by the Anderson-Darling test, to evaluate the normality of each component. 

To identify the number of joint and individual clusters, we use the fact that the total rank of the cluster structure in the concatenated matrix, $X$, is given by
\[E = r + r_{1}+\cdots + r_{M},\]
and the rank of the cluster structure in the original data $X_{m}$ is $E_{m} = r + r_{m}$ for $m=1,\dots,M$. As the number of clusters is given by $r+1$ and $r_{m}+1$ respectively, we can determine $E$ and $E_{1},\dots, E_{M}$ in the data and use them to calculate $K$ and $K_{1}, \dots, K_{M}$. We follow the two step procedure:
\begin{enumerate}
	\item Estimate the number of relevant subspaces $E$ in $X$, when the ranks of the individual structures are fixed to zero: test the normality of the $i$th joint component scores for increasing $i$, until the last non-normally distributed component is found and set $E$ to the component number. 
	\item Estimate the number of relevant subspaces $E_{m}$ in 
	 $X_{m}$: For each $m=1,\dots, M$, test the normality of the $i$th component scores for increasing $i$, until the last non-normally distributed component is found and set $E_{m}$ to the component number. 
\end{enumerate}
Now, the number of joint clusters  is given as
\begin{equation}
K = \frac{E_{1}+ \dots+ E_{M}-E}{M-1} + 1, \label{clusterEq}
\end{equation}
while the number of individual clusters is given as $K_{m} = E_{m} - K + 2$ for $m=1,\dots,M$. 

\section{Simulations}\label{simulation}
We compare JIC to the iCluster procedure by simulating two different settings; only joint clusters and both joint and data type-specific clusters. In both settings, three different data types are integrated, $M=3$, and the number of clusters is first assumed known, then estimated by the procedure described in Section \ref{findingK}. 

\begin{table}
\centering
\begin{tabular}{r|c|cccc}
 &iCluster & JIC: joint & $X_{1}$ & $X_{2}$ & $X_{3}$ \\ \hline \hline
 Setting I: Precision  &  0.998& 0.985  & -& -& - \\ 
Correctly estimated $K$    &                & 97\%  & 96\% & 95\% & 98\%     \\ \hline \hline
Setting II: Precision &  0.415   & 0.933 & 0.950  & 0.791  &0.874 \\
Correctly estimated $K$    &   &  89\%    &90\% &88\% & 88\% 
\end{tabular}
\caption{Mean precision of estimated cluster assignment (over 100 simulations), when the numbers of clusters are known. Percentage of times the numbers of clusters are correctly estimated.  
\label{accuracy}}
\end{table}

\subsection{Setting I: Joint cluster structure}
First, we simulate 5 joint clusters, present in all three data sets. Specifically, $n=150$, where $j=1,\dots,30$ belongs to the first cluster, $j=31,\dots,60$ belongs to the second cluster and so on, giving 30 observations in each cluster. The joint latent variable $Z_{J}^{T}$, with the indicator vectors as columns, is an $n \times 4$ matrix
\[ Z_{J}^{T} = 
\begin{bmatrix} 	1 &0 &\cdots  \\ \vdots & \vdots & \\	0 & 1 &\cdots \\	\vdots & \vdots &  \\ 0 &0&\cdots \end{bmatrix}  
.\]
Each row contains a single '1' indicating the assignment of the observation to the cluster corresponding to the column number. The last cluster is, however, specified by only zeros. The loading matrices $W_{1}, W_{2}$ and $W_{3}$ are of the same dimension $200 \times 4$ ($p_{1}=p_{2}=p_{3}=200$). We generate the loadings according to a standard normal distribution and normalize the matrices, such that $W_{m}^{T}W_{m}= I$ for $m=1,2,3$. Within each $W_{i}$, the columns are also made orthogonal to each other. The three data sets are generated by 
\begin{align*}
X_{1} &= c W_{1}^{T}Z_{J} + \eps_{1}, \\
X_{2} &= c W_{2}^{T}Z_{J} + \eps_{2}, \\
X_{3} &= c W_{3}^{T}Z_{J} + \eps_{3}, 
\end{align*}
with standard normally distributed errors, $\eps_{m} \sim N(0,I)$, and $c=80$. 

In the simulation, we first assume $K=5$ known and compare the estimated cluster assignments to the true clusters in terms of the precision. Secondly, we assume $K$ unknown and estimate it by the procedure in Section \ref{findingK}. Under Setting I in Table \ref{accuracy}, the precision of JIC compared to the iCluster methodology is shown. We see that iCluster and JIC perform equally well in the situation with only joint clusters. In the case of unknown number of clusters, $K$ was correctly estimated in 97\% of the simulated cases, as seen in Table \ref{accuracy}.

\subsection{Setting II: Joint and individual clusters}
In the second setting, two data type-specific clusters are added in each of the three data sets. The observations are randomly assigned to one of two clusters, such that the data type-specific latent variables $Z_{1}, Z_{2}$ and $Z_{3}$ are vectors with random ones and zeros. For the loadings matrices $V_{1},V_{2}$ and $V_{3}$ of dimension $200 \times 1$, the loadings are randomly generated according to a standard normal distribution and normalized, such that $V_{m}^{T}V_{m}=1$ for $m=1,2,3$. 

To obtain an identifiable decomposition, each $Z_{m}$ is made orthogonal to the columns of $Z_{J}$. The three data sets are generated by the model
\begin{align*}
X_{1} &= c W_{1}^{T}Z_{J} + c_{1}V_{1}^{T}Z_{1} + \eps_{1}, \\
X_{2} &= c W_{2}^{T}Z_{J} + c_{2}V_{2}^{T}Z_{2} + \eps_{2}, \\
X_{3} &= c W_{3}^{T}Z_{J} + c_{3}V_{3}^{T}Z_{3} + \eps_{3}, 
\end{align*}
with standard normally distributed noise, $\eps_{m} \sim N(0,I)$, $c=80$ and $c_{1}=c_{2}=c_{3}=30$. First, the correct numbers of clusters, $K=5$ and $K_{1} = K_{2}=K_{3}=2$,
are assumed known and the joint and individual clustering are compared to the true cluster memberships. The precisions are shown in Table \ref{accuracy} under Setting II. For iCluster, only the precision of the joint clustering is displayed. 

We see that JIC is highly superior to the iCluster method in recovering the joint cluster as the individual clusters clearly obscure the joint signal. We also see that JIC performs well with a high precision for both the joint and individual clusters. Table \ref{accuracy} shows that when $K, K_{1},K_{2}$ and $K_{3}$ are assumed unknown, they can be correctly estimated by the procedure in Section \ref{findingK}.

\section{Example: the Metabric study}\label{metabric}
To illustrate JIC, we will analyze the data from the Metabric study \citep{curtis2012genomic} with a discovery set consisting of the gene expression and somatic copy number aberrations (CNAs) of 997 breast cancer tumor samples. The data are available through European Genome-Phenome Archive (http://www.ebi.ac.uk/ega/), under accession number EGAS00000000083. For the analysis, we select the 1000 genes and CNA locations with the largest variability. The CNAs are considered gene locations with tumor-specific differences in copy number compared to a healthy control, and are recorded as the count of gene copies, transformed to a log2 scale. Also recorded is disease-specific survival, together with the clinical variables: age, estrogen status, treatment and PAM50 classification. The outline of the analysis is as follows: First, the number of joint and individual clusters is chosen. Then, both clusterings are tested for differences in survival time and explored with regard to the available clinical variables.  

%

\begin{figure}%
\centering \includegraphics[width=\columnwidth]{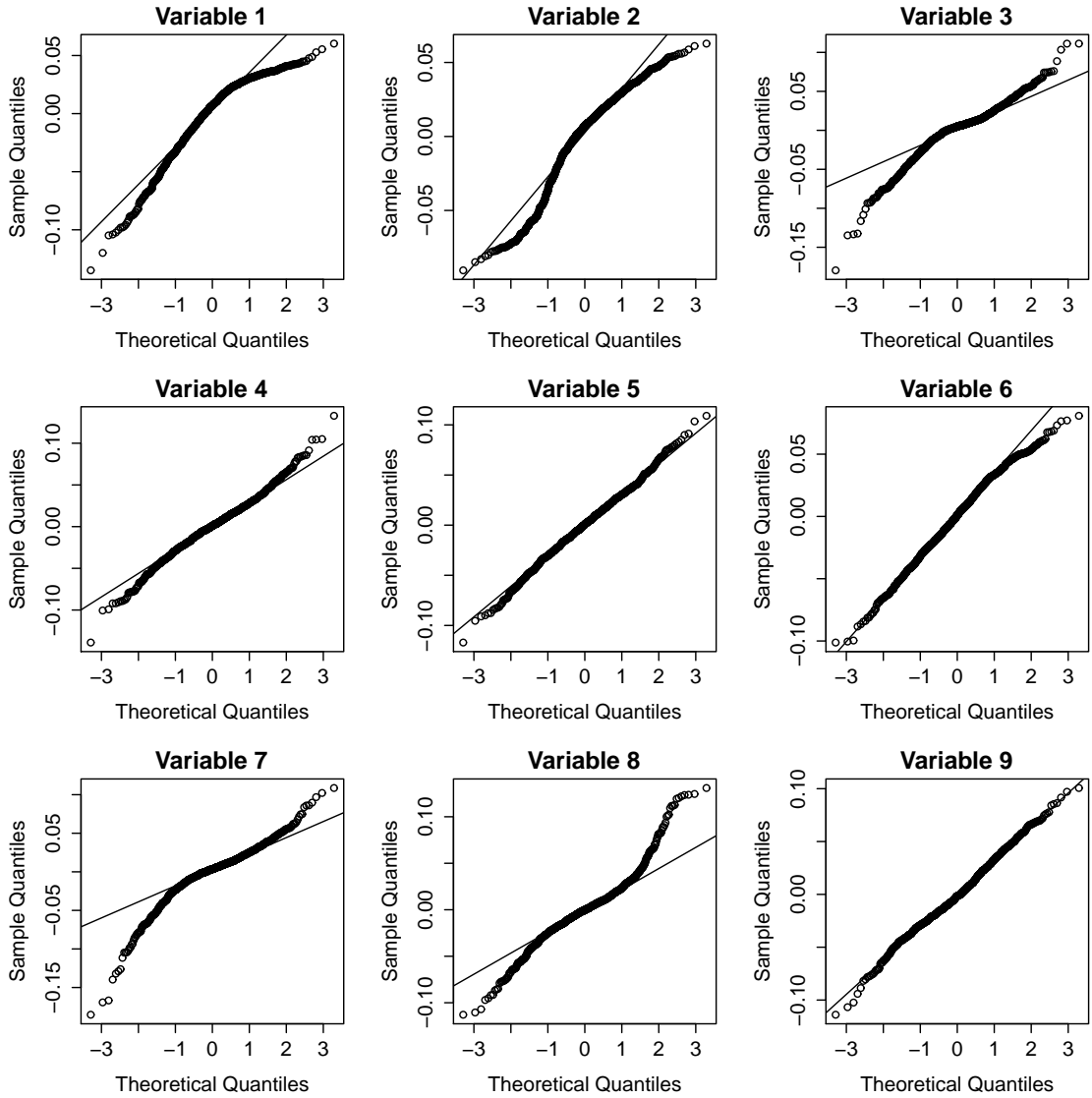}%
\caption{Normal quantile-quantile plots for the first 9 joint component scores. The 5th and 9th do not exhibit clear deviations from normality.}%
\label{NormalJoint}%
\end{figure}

We determine the number of joint, expression-specific and CNA-specific clusters, $K,K_{1},K_{2}$ according to the procedure described in Section \ref{findingK}. Figure \ref{NormalJoint} displays the qq-plots of the first 9 joint component scores, not allowing for individual structures. Generally, it is seen that the component scores are closer to being normally distributed as the component number increases. The first, second, third and fourth joint components are clearly not normally distributed, while the 5th and 6th are borderline cases. Then, again the 7th and 8th component scores clearly deviate from normality, while the 9th component does not seem to deviate significantly. This is confirmed by the Anderson-Darling test, and we therefore determine the rank of the complete joint and individual cluster structure to be $E=8$. It would also be possible stop at the fifth component, but with an exploratory aim of the analysis and the clear signs of structure in the 7th and 8th component in mind, we choose to include more components. 

We examine the qq-plots of the first three component scores of the original expression data. This shows that the first component is clearly non-normal, while the second component is a borderline case and the third component does not deviate significantly from normality.  We therefore determine the number of relevant subspaces in the expression data to be $E_{1}=2$. We also examine the qq-plots of the first 8 component scores of the original CNA data. However, when analyzing the CNA data individually, the assumption of normally distributed noise is not properly fulfilled due to the discrete nature of the copy number counts . All of the qq-plots therefore show a clear deviation from normality, and as the total rank of the original data cannot exceed $E$, we set $E_{2}=8$. 

\begin{figure}
\centering
\includegraphics[width=\columnwidth]{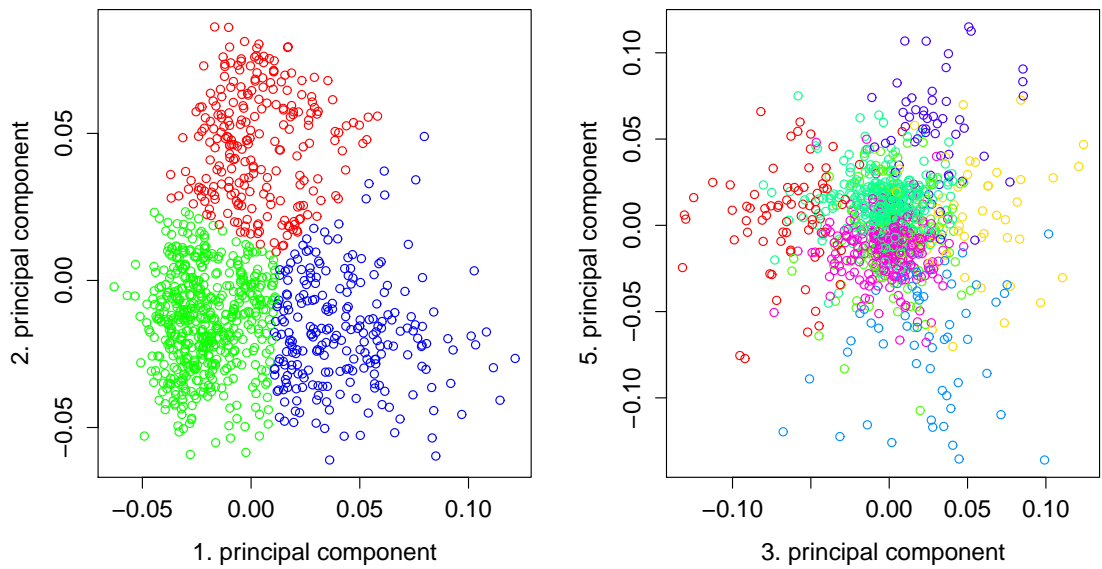}%
\caption{a) The 1. and 2. joint component scores with the three joint clusters in different coloring. b) The 3. and 5. CNA-specific component scores with the seven CNA clusters in different coloring.}%
\label{jointCluster}%
\end{figure}

With $E=8, E_{1}=2, E_{2}=8$, we calculate the number of clusters using  \eqref{clusterEq}: 
\[K=3, \quad K_{1}=1, \quad K_{2}=7,\]
meaning we use three joint clusters, no expression-specific clusters and seven   CNA-specific clusters. Figure \ref{jointCluster}a) displays the first and second joint component scores, and we see that the first component discriminates between the 'purple' and 'green' cluster, while the second component separates out the 'red' cluster. Comparing the clusters in terms of clinical covariates, reveals that the 'red' cluster coincide with the Estrogen Receptor (ER) status of the patients, as most ER-negative patient cases are present in the 'red' cluster. Within the PAM50 classification, ER-negative cases are mainly of Basal or HER2-type, meaning the 'red' cluster mainly consists of these two cancer subtypes, as observed in Table \ref{JointPam50}.

\begin{figure}
\centering\includegraphics[width=\columnwidth]{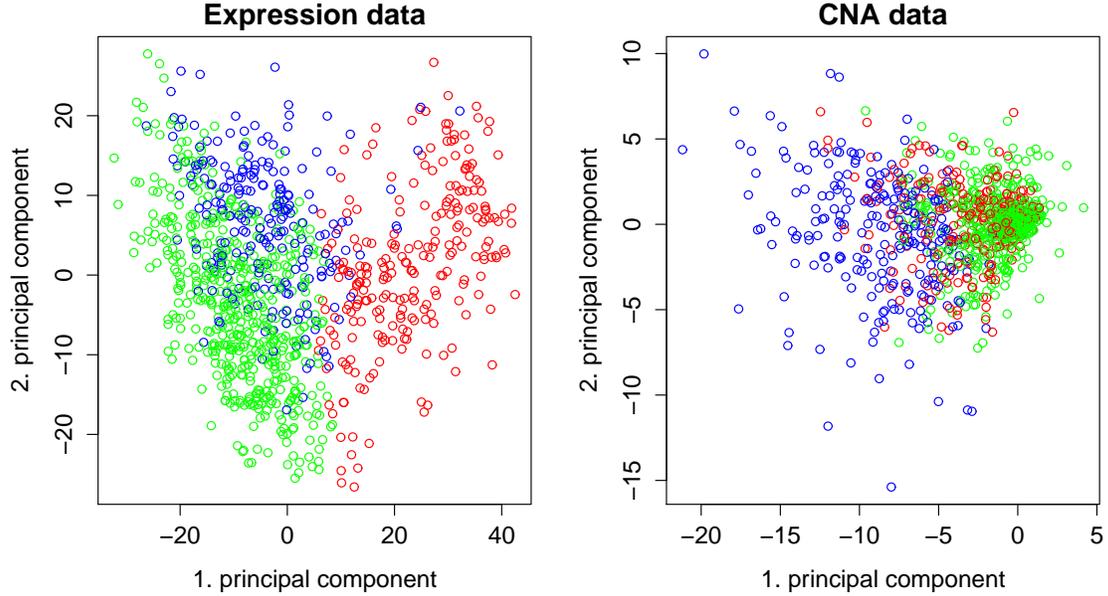}%
\caption{The first and second principal component of the original expression data and the copy number aberrations data, colored with the three joint clusters.}%
\label{originalCluster}%
\end{figure}

To investigate the relationship between the joint clusters and the original data, Figure \ref{originalCluster} displays the first and second principal component scores of the original expression and CNA data with the coloring of the joint clusters. For the expression data, it is clear that the main differences are between the 'red' cluster and the two other clusters. In the CNA data, on the other hand, the observations in the 'red' cluster are randomly scattered, while the two other clusters are quite distinct. 

To visualize the seven CNA-specific clusters, we look at the 3rd and 5th component scores, as seen in Figure \ref{jointCluster}b). For the Figure, it is seen that the 3rd component distinguish between the 'yellow' and 'red' cluster, while the 5th shows the difference between the 'light blue' and 'purple' group. It is also observed that the remaining three clusters, especially the 'green' and 'lilac', are neutral groups situated at the origin.

\subsection{Connections with survival, Metabric- and PAM50 classification}
The joint and CNA-specific clusters are independently evaluated with regard to survival through Kaplan-Meier estimates. When comparing the three joint clusters against each other and the seven CNA-specific clusters against each other, both clusterings were shown to give significant differences by the logrank test ($p = 8.7\cdot 10^{-7}$  and $p =1.8\cdot 10^{-7}$ for joint and CNA clusters, respectively). Also, when adjusting for age and treatment in a Cox proportional hazards model, both the joint and CNA-specific clusters are significant ($p=0.02$ and $p=0.0004$, respectively) by the likelihood ratio test. 

\begin{figure}
\centering\includegraphics[width=\columnwidth]{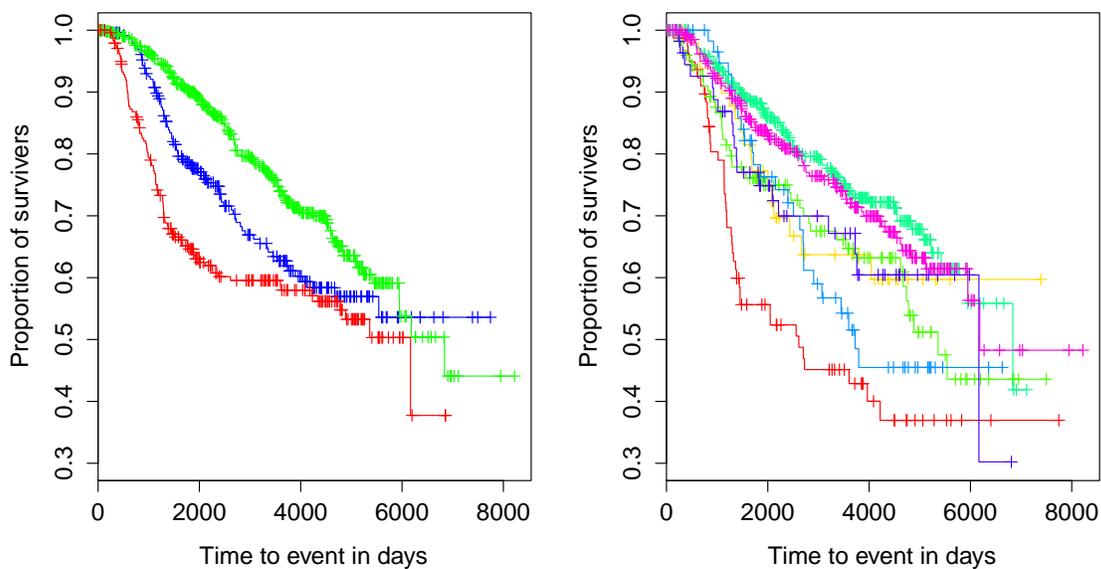}%
\caption{a) A Kaplan-Meier survival plot of the 3 joint clusters. b) A  Kaplan-Meier survival plot of the 7 CNA clusters.}%
\label{survival}%
\end{figure}

Figure \ref{survival}a) displays the Kaplan-Meier plot of the three joint clusters, revealing the 'red' cluster to be a high mortality risk group, the 'purple' cluster to be an intermediate risk group and the 'green' cluster to be a low risk group. Figure \ref{survival}b) displays the Kaplan-Meier plot for the seven clusters only present in the CNA data. Interestingly, the two neutral 'dark green' and 'lilac' clusters, situated at the origin of Figure \ref{jointCluster}b), are low-risk mortality groups. These exhibit few somatic changes in the overall copy number patterns compared to healthy tissue. Conversely, the 'red','blue', 'purple' and 'yellow' groups with quite specific aberration patterns, all exhibit an increased risk of mortality. Especially, the copy number aberrations associated with a negative 3rd component in CNA structure results in highly increased risk, compared to the other groups. 

\begin{table}\centering
\begin{tabular}{cccccc}
  \hline
 Risk& Basal & Her2 & LumA & LumB &  Normal \\ 
  \hline
  High & 115 &  63 &  &  &    37 \\ 
  Low  &  &  & 390 & 100 &    \\ 
  Intermediate &  &  &  63 & 152 & \\\hline    
Total &118&86&456&268&58
\end{tabular}
\caption{The distribution of patients from the PAM classification in the three joint clusters. For clarity, entries constituting less than 10\% row-wise are not shown.  
}\label{JointPam50}
\end{table}

\begin{table}\centering
\begin{tabular}{cccccc}
  \hline
Risk & Basal & Her2 & LumA & LumB &  Normal \\ 
  \hline
	Very high (red)  &  &  31 &  13 &  29 &   \\ 
  High (yellow) &  &   6 &   7 &  35 &    \\ 
  High (light blue) &  &  &  31 &  25 &    \\   
	High (purple) &  &  &  23 &  26 &    \\ 
	High (lime)&  19 &  &  56 &  37 &    \\ 
  Low (green) &  51 &  & 184 &  69 &    \\ 
  Low (pink) &  36 &  & 151 &  47 &  \\   \hline 
	Total &118&86&456&268&58
\end{tabular}
\caption{The distribution of patients from the PAM classification in the seven individual clusters. For clarity, entries constituting less than 10\% row-wise are not shown.  
}\label{IndividPam50}
\end{table}


The clusters found by JIC are related to the PAM50 classification \citep{perou2000molecular} and the 10 breast cancer subgroups identified by the initial Metabric study \citet{curtis2012genomic}. The Tables \ref{JointPam50}-\ref{IndividMetabric} display the distribution of patients according to the different clusterings. 

Table \ref{JointPam50} displays the agreement between the three joint clusters and five subtypes in the PAM50 classification, and it is clear that the high risk cluster consists of Basal, Her2 and Normal-type tumors, while the low and intermediate are dominated by Luminal A and B. The low risk group has a majority of Luminal A cases, while the intermediate group has a majority of Luminal B cases. Table \ref{IndividPam50} displays the agreement between the seven CNA clusters and PAM50, but we observe no clear patterns here. An interesting observation is that the Basal and Her2 cases do not belong to the same cluster, indicating that the two classes differ in specific copy number alterations as also suggested by the Metabric study \citep{curtis2012genomic}. The Her2 group is mainly found in the very high risk 'red' group. The Luminal A and B cases are evenly distributed among all the clusters, but with a pivot in the two low risk groups. 

\begin{table}\centering
\begin{tabular}{rcccccccccc}
  \hline
 Risk& 1 & 2 & 3 & 4 & 5 & 6 & 7 & 8 & 9 & 10 \\ \hline
  High &  &  &  &  68 &  50 &  &  &  &  &  87 \\ 
  Low &  &  & 150 &  95 &  &  &  68 & 127 &  &  \\ 
  Intermediate &  64 &  &  &  &  &  32 &  38 &  &  44 &  \\\hline 
		Total &75 & 45 &155 &167  &94 & 44 &109 &143 & 67&  96 
\end{tabular}
\caption{The distribution of patients from the ten Metabric clusters \citep{curtis2012genomic} in the three joint clusters. For clarity, entries constituting less than 10\% row-wise are not shown. 
}\label{JointMetabric}
\end{table}

\begin{table}\centering
\begin{tabular}{rcccccccccc}
  \hline
Risk & 1 & 2 & 3 & 4 & 5 & 6 & 7 & 8 & 9 & 10 \\ 
  \hline
  Very high (red) &  &  &  &  &  69 &  &  &  &  &  \\ 
  High (yellow) &  38 &  &  &  &  &  &  &  &  &  \\ 
  High (light blue) &  &  40 &  &  &  &  &  &  &  &  \\ 
  High (purple) &  &  &  &  &  &  34 &  &  &  &  \\ 
	High (lime) &  &  &  &  &  &  &  33 &  19 &  29 &  18 \\ 
  Low (green) &  &  &  76 &  81 &  &  &  &  66 &  &  39 \\  
	Low (pink) &  &  &  61 &  64 &  &  &  37 &  49 &  &  31  \\ \hline 
			Total &75 & 45 &155 &167  &94 & 44 &109 &143 & 67&  96 
\end{tabular}
\caption{The distribution of patients from the ten Metabric clusters \citep{curtis2012genomic} in the seven individual clusters. For clarity, entries constituting less than 10\% row-wise are not shown.  
}\label{IndividMetabric}
\end{table}

Table \ref{JointMetabric} shows the distribution of patients between the 10 integrative Metabric clusters found by \citet{curtis2012genomic} and the three joint clusters. Here we observe that the high risk group mainly consists of the Metabric cluster 10, 4 and 5, where the 10th subgroup largely corresponds to the Basal subtype in the PAM50 classification. Further the low risk group consists mainly of  Metabric clusters 3 and 8, together with 4 and 7. The intermediate risk group is less clear, but corresponds largely to Metabric clusters 1, 6 and 9. 

Table \ref{IndividMetabric} displays the distinct pattern of the correspondence between the ten  Metabric clusters and the seven CNA-specific clusters found by JIC. The four groups with the highest risk profile corresponds uniquely to four Metabric clusters: The very high risk 'red' group corresponds to the 5th cluster, the high risk 'yellow' group to the 1st cluster, the high risk 'light blue' group to the 2nd cluster and the high risk 'purple' group to the 6th Metabric cluster. The 9th Metabric cluster is only found as a part of the high risk 'lime' group, while the remaining Metabric clusters 3,4,7,8 and 10 are evenly distributed between the high risk 'lime' group and the two low risk groups. 

In conclusion, these observations suggest that there are two independent mechanisms influencing patient survival. From the PAM50 classification, there is a substantial mortality risk difference between the Basal and Her2 on one side and the Luminal A and B on the other. This seems to be the main driver of survival differences, but specific copy number alterations will in addition have an effect. This is seen from the highest risk CNA-specific cluster, which contains a large degree of Luminal A and B (Table \ref{IndividPam50}), but only the 5th Metabric cluster (Table \ref{IndividMetabric}). There exist certain  copy number aberrations, which override the overall group differences between the Basal/Her2 and the Luminal subtypes. The same reasoning also applies to the other high risk CNA-specific clusters.

\section{Discussion}

%
%

The Joint and Individual Clustering (JIC) contributes to the increased need for integrative procedures within genomics, by decomposing patient samples into joint and individual clusters simultaneously. This improves the  understanding of cancer subtypes across genetic data types, as completely independent clusterings can both explain significant differences in survival. This suggests that in addition to clusters of cancer subtypes, found jointly in different data types, there exists, in for instance CNA data, independent groups related to other clinical variables, possibly age, smoking or other environmental influences. The results also agree with earlier analysis of the Metabric data by \citet{curtis2012genomic}, where the iCluster method was used to identify 10 joint clusters. Specifically, four of the seven CNA-specific clusters correspond exactly to four of the joint clusters found by \citet{curtis2012genomic}, suggesting that these are not joint clusters, but instead  specific for the CNA data.  

The crucial step of how to select the number of clusters proved to be difficult in our setting due to the high-dimensionality of the data. The use of cluster separation measures or cluster reproducibility by sub-sampling did not yield good results within JIC and therefore the more subjective normality-based approach was used. The selection of the number of clusters will always contain subjective aspects, and our selection procedure makes these choices particularly transparent. 

\section*{Acknowledgment}
The authors would like to thank J.S. Marron for pointing out certain of the inner workings of the JIVE algorithm. Funding for the project was provided by  the Norwegian Cancer society under award 744088. 
\bibliographystyle{chicago}
\bibliography{biblioInteg1}


\end{document}